\pdfoutput=1
\documentclass[aps,prd,preprint,superscriptaddress,tightenlines,nofootinbib]{revtex4}

\long\def\symbolfootnote[#1]#2{\begingroup%
\def\thefootnote{\fnsymbol{footnote}}\footnote[#1]{#2}\endgroup}

\usepackage{amsfonts,amsmath,amsthm,amssymb}
\usepackage{mathrsfs}
\usepackage{bm}
\usepackage{color}
\usepackage{epsfig}

\newcommand{\PRE}[1]{{#1}}   

\newcommand{\beq}{\begin{equation}}
\newcommand{\eeq}{\end{equation}}
\newcommand{\bea}{\begin{flushleft} \begin{eqnarray}}
\newcommand{\eea}{\end{eqnarray}\end{flushleft}}

\newcommand{\comment}[1]{}

\newcommand{\el}[1]{\label{#1}}
\newcommand{\er}[1]{\eqref{#1}}

\newcommand{\ci}[1]{}

\newcommand{\ke}{\rangle}
\newcommand{\br}{\langle}

\newcommand{\p}{\partial}

\newcommand{\ba}{\begin{eqnarray}}
\newcommand{\ea}{\end{eqnarray}}
\newcommand{\be}{\begin{equation}}
\newcommand{\ee}{\end{equation}}
\newcommand{\bay}[1]{\left(\begin{array}{#1}}
\newcommand{\eay}{\end{array}\right)}

\def\xt{{\theta}}

\def\CD{{\cal D}}

\def\CR{{\cal R}}

\usepackage[usenames,dvipsnames]{xcolor}
\definecolor{orange}{cmyk}{0,0.5,1,0}
\definecolor{rossoCP3}{cmyk}{0,.88,.77,.40}
\definecolor{graa}{rgb}{0.8,0.8,0.8}
\definecolor{blaa}{rgb}{0.2,0.2,0.6}

\begin{document}

\preprint{
\hfil
\begin{minipage}[t]{3in}
\begin{flushright}
\vspace*{.4in}
MPP--2015--155\\
LMU-ASC 45/15\\
\end{flushright}
\end{minipage}
}

\title{\PRE{\vspace*{0.9in}} \color{rossoCP3}{Stringy origin of 
   diboson and dijet excesses at the LHC
}}

\author{{\bf Luis A. Anchordoqui}}

\affiliation{Department of Physics and Astronomy,\\  Lehman College, City University of
  New York, NY 10468, USA
\PRE{\vspace*{.05in}}
}

\affiliation{Department of Physics,\\
 Graduate Center, City University
  of New York,  NY 10016, USA
\PRE{\vspace*{.05in}}
}

\affiliation{Department of Astrophysics,\\
 American Museum of Natural History, NY
 10024, USA
\PRE{\vspace*{.05in}}
}

\author{{\bf Ignatios Antoniadis}}
\affiliation{LPTHE, UMR CNRS 7589\\
Sorbonne Universit\'es, UPMC Paris 6, 75005 Paris, France
\PRE{\vspace*{.05in}}}

\affiliation{Albert Einstein Center, Institute for Theoretical Physics\\
Bern University, Sidlerstrasse 5, CH-3012 Bern, Switzerland
\PRE{\vspace*{.05in}}}

\author{{\bf Haim \nolinebreak Goldberg}}
\affiliation{Department of Physics,\\
Northeastern University, Boston, MA 02115, USA
\PRE{\vspace*{.05in}}
}

\author{{\bf Xing Huang}}
\affiliation{Department of Physics, \\
National Taiwan Normal University, Taipei, 116, Taiwan
\PRE{\vspace*{.05in}}
}

\author{{\bf Dieter L\"ust}}

\affiliation{Max--Planck--Institut f\"ur Physik, \\ 
 Werner--Heisenberg--Institut,
80805 M\"unchen, Germany
\PRE{\vspace*{.05in}}
}

\affiliation{Arnold Sommerfeld Center for Theoretical Physics 
Ludwig-Maximilians-Universit\"at M\"unchen,
80333 M\"unchen, Germany
\PRE{\vspace{.05in}}
}

\author{{\bf Tomasz R. Taylor}}

\affiliation{Department of Physics,\\
 Northeastern University, Boston, MA 02115, USA 
 \PRE{\vspace*{.05in}}
}

\PRE{\vspace*{.15in}}

\begin{abstract}\vskip 2mm
  \noindent 
  Very recently, the ATLAS and CMS collaborations reported diboson and
  dijet excesses above standard model expectations in the invariant
  mass region of $1.8 -2.0~{\rm TeV}$. Interpreting the diboson excess
  of events in a model independent fashion suggests that the vector
  boson pair production searches are best described by $WZ$ or $ZZ$
  topologies, because states decaying into $W^+W^-$ pairs are strongly
  constrained by semileptonic searches.  Under the assumption of a low
  string scale, we show that both the diboson and dijet excesses can
  be steered by an anomalous $U(1)$ field with very small coupling to
  leptons.  The Drell-Yan bounds are then readily avoided because of
  the leptophobic nature of the massive $Z'$ gauge boson. The
  non-negligible decay into $ZZ$ required to accommodate the data is a
  characteristic footprint of intersecting D-brane models, wherein the
  Landau-Yang theorem can be evaded by anomaly-induced operators
  involving a longitudinal $Z$. The model presented herein can be
  viewed purely field-theoretically, although it is particularly well
  motivated from string theory.  Should the excesses become
  statistically significant at the LHC13, the associated $Z\gamma$
  topology would become a signature consistent only with a stringy
  origin.
\end{abstract}

\maketitle

Very recently, searches for narrow resonances at the ATLAS and CMS
experiments uncovered various peaks in invariant mass distributions near
$2~{\rm TeV}$: {\it (i)}~The ATLAS search for diboson production
contains a $3.4\sigma$ excess at $\sim 2~{\rm TeV}$ in boosted jets of
$WZ$~\cite{Aad:2015owa}. The global significance of the discrepancy
above standard model (SM) expectation is $2.5\sigma$. The invariant
mass range with significance above $2\sigma$ is $\sim 1.9$ to
$2.1~{\rm TeV}$. Because the search is fully hadronic, the capability
for distinguishing gauge bosons is narrowed. Therefore, many of the
events can also be explained by a $ZZ$ or $WW$ resonance, yielding
excesses of $2.9\sigma$ and $2.6\sigma$ in these channels
respectively. {\it (ii)}~The CMS search for diboson production
(without distinguishing between the $W-$ and $Z-$tagged jets) has a
1.4$\sigma$ excess at $\sim 1.9~{\rm TeV}$~\cite{Khachatryan:2014hpa},
and the search for diboson production with a leptonically tagged $Z$
yields a $1.5\sigma$ excess at invariant mass $\sim 1.8~{\rm
  TeV}$~\cite{Khachatryan:2014gha}.  {\it (iii)}~The CMS search for
dijet resonances finds a $2.2\sigma$ excess near
1.8~TeV~\cite{Aad:2014aqa}. {\it (iv)}~Around the same invariant mass
ATLAS also recorded an excess in the dijet distribution with a
$1\sigma$ significance~\cite{Khachatryan:2015sja}. {\it (v)} The CMS
search for resonant $HW$ production yields a $2.1\sigma$ excess in the
energy bin of 1.8 to $1.9~{\rm TeV}$; here the Higgs boson is highly
boosted and decays into $b \bar b$, whereas the $W$ decays into
charged leptons and neutrinos~\cite{CMS-14-010}. Barring the three
ATLAS analyses in diboson production, all these excesses are
completely independent. 

Although none of the excesses is statistical significant yet, it is
interesting to entertain the possibility that they correspond to a
real new physics signal. On this basis, with the assumption that
all resonant channels are consistent with a single resonance energy, a model
free analysis of the various excesses has been recently
presented~\cite{Brehmer:2015cia}.  The required cross sections to
accommodate the data are quite similar  for $WZ$ and
$ZZ$ final states, which can be considered as roughly the same measurement.
A pure $WW$ signal is disfavored and could only describe the data in
combination with another signal. This is because the CMS single lepton
analysis sets an upper bound of 6.0~fb at 95\% C.L.~\cite{Khachatryan:2014gha}  and a cross
section of this magnitude is needed to reproduce the hadronic
excesses. Moreover, the CMS dilepton search has a small excess that
this channel cannot explain~\cite{Khachatryan:2014gha}.

Several explanations have been proposed to explain the excesses
including a new charged massive spin-1 particle coupled to the
electroweak sector (which can restore the left-right
symmetry)~\cite{Hisano:2015gna}, strong dynamics engendering composite
models of the bosons~\cite{Fukano:2015hga},  dark matter annihilation
into right-handed fermions~\cite{Alves:2015mua}, a resonant triboson
simulating a diboson through judicious choice of
cuts~\cite{Aguilar-Saavedra:2015rna}, and a heavy scalar~\cite{Chen:2015xql}.  In this Letter we adopt an
alternate path. We assume that the source of the excesses originates
in the decay of a new abelian gauge boson that suffers a mixed anomaly
with the SM, but is made self-consistent by  the Green-Schwarz (GS)
mechanism~\cite{Green:1984sg}. Such gauge bosons occur naturally in D-brane TeV-scale
string compactifications~\cite{Antoniadis:1998ig}, in which the gauge
fields are localized on D-branes wrapping certain compact cycles on an
underlying geometry, whose intersection can give rise to chiral
fermions~\cite{Blumenhagen:2005mu}. The SM arises from strings
stretching between D-branes which belong to the ``visible'' sector.
Additional D-branes are generally required to cancel RR-tadpoles, or
to ensure that all space-filling charges cancel. These additional
D-branes generate gauge groups beyond the SM which forge the
``hidden'' sector. 

There are two unrivaled phenomenological ramifications for intersecting
D-brane models: the emergence of Regge excitations at parton collision
energies $\sqrt{\hat s} \sim$ string scale $\equiv M_s;$ and the
presence of one or more additional $U(1)$ gauge symmetries, beyond the
$U(1)_Y$ of the SM. The latter derives from the property that, for $N
> 2$, the gauge theory for open strings terminating on a stack of $N$
identical D-branes is $U(N)$ rather than $SU(N)$. (For $N = 2$ the
gauge group can be $Sp(1) \cong SU(2)$ rather than $U(2)$.)  In a
series of recent publications we have exploited both these
ramifications to explore and anticipate new-physics signals that could
potentially be revealed at the LHC. Regge excitations most distinctly
manifest in the $\gamma +$ jet~\cite{Anchordoqui:2007da} and
dijet~\cite{Lust:2008qc} spectra resulting from their decay. The extra
$U(1)$ gauge symmetries beyond hypercharge have (in general) triangle
anomalies, but are cancelled by the GS mechanism and the $U(1)$ gauge
bosons get St\"uckelberg masses. We have used a minimal
D-brane construct to show that the massive $U(1)$ field, the $Z'$,
can be tagged at the LHC by its characteristic decay to dijets or
dileptons~\cite{Anchordoqui:2011ag}. In the framework of this model
herein we  adjust the coupling strengths to be simultaneously
consistent with the observed dijet excess and the lack of a
significant dilepton excess.  Concurrently we show that the model is
also consistent with the ATLAS diboson excess as it allows for
production of $Z$-pairs. At the level of effective Lagrangian, the
operator contributing to the $Z'ZZ$ amplitude is induced by the GS
anomaly cancellation.
 
In our calculations we will adopt as
benchmarks:
\begin{align} 
& \sigma (pp \to Z') \times {\cal  B} (Z' \to ZZ/WW) \sim
5.5^{+5.1}_{-3.7}~{\rm fb}  &~\text{\cite{Brehmer:2015cia}},  \label{uno}\\
& \sigma(pp \to Z') \times {\cal B} (Z' \to jj)   \sim
91^{+53}_{-45}~{\rm fb} &~\text{\cite{Brehmer:2015cia}}, \label{dos} \\  
& \sigma(pp \to Z') \times {\cal B} (Z' \to e^+ e^-)   <   0.2~{\rm fb}~(95\%~{\rm C.L.}) &\text{\cite{Aad:2014cka}}, \label{cuatro} \\
& \sigma(pp \to Z') \times {\cal B} (Z'\to HZ)  < 12.9~{\rm
  fb}~(95\%~{\rm C.L.}) &~~\text{\cite{Khachatryan:2015bma}}. \label{cinco}  
\end{align}

To develop our program in the simplest way, we will work within the
construct of a minimal model with 4 stacks of
D-branes in the visible sector. The basic setting of the gauge theory
is given by $U(3)_a \times Sp(1)_b \times U(1)_c \times U(1)_d$~\cite{Cremades:2003qj}.  The
LHC collisions take place on the (color) $U(3)_a$ stack of
D-branes. In the bosonic sector the open strings terminating on this
stack contain, in addition to the $SU(3)_C$ octet of gluons $g_\mu^a$,
an extra $U(1)$ boson $C_\mu$, most simply the manifestation of a
gauged baryon number. The $Sp(1)_b$ stack is a terminus for the
$SU(2)_L$ gauge bosons $W^a_\mu$. The $U(1)_Y$ boson $Y_\mu$ that
gauges the usual electroweak hypercharge symmetry is a linear
combination of $C_\mu$ and the $U(1)$ bosons $B_\mu$ and $X_\mu$
terminating on the separate $U(1)_c$ and $U(1)_d$ branes.  Any vector
boson orthogonal to the hypercharge, must grow a mass so as
 to avoid long range forces between baryons other than gravity
and Coulomb forces. The anomalous mass growth allows the survival of
global baryon number conservation, preventing fast proton decay~\cite{Ghilencea:2002da}.

The content of the hypercharge operator is  given by
\begin{equation}
Q_Y = \frac{1}{6} Q_a - \frac{1}{2} Q_c  + \frac{1}{2} Q_d  \, .
\label{hyperchargeY}
\end{equation}
We also extend the fermion sector by including the right-handed
neutrino, with $U(1)$ charges $Q_a = 0$ and $Q_{c}= Q_{d} = -1$. The
chiral fermion charges of the model are summarized in
Table~\ref{table}. It is straightforward to see that the chiral
multiplets yield a $[U(1)_a SU(2)_L^2]$ mixed anomaly through triangle
diagrams with  fermions running in the loop. This anomaly is
cancelled by the GS mechanism, wherein closed string couplings yield
classical gauge-variant terms whose gauge variation cancels the
anomalous triangle diagrams. The extra abelian gauge field becomes
massive by the GS anomaly cancellation, behaving at low energies as a
$Z'$ with a mass in general lower than the string scale by an order of
magnitude corresponding to a loop factor.  Even though the divergences
and anomaly are cancelled, the triangle diagrams contribute an
univocal finite piece to an effective vertex operator for an
interaction between the $Z'$ and two $SU(2)_L$ vector
bosons~\cite{Anastasopoulos:2006cz}. This is a distinguishing aspect
of the D-brane effective theory, which features a noticeable decay
width of the $Z'$ into $WW$, $ZZ$, and $Z\gamma$.\footnote{The
  Landau-Yang theorem~\cite{Landau:1948kw}, which is based on simple
  symmetry arguments, forbids decays of a spin-1 particle into two
  photons.}

\begin{table}
  \caption{Chiral fermion spectrum of the D-brane model.}
\begin{center}
\begin{tabular}{ccccccc}
\hline
\hline
~~~Fields~~~ & ~~~Sector~~~  & ~~~Representation~~~ & ~~~$Q_a$~~~ & ~~~$Q_d$~~~ & ~~~$Q_c$~~~ & ~~~$Q_Y$~~~ \\
\hline
 $U_R$ &   $\phantom{^*}a \leftrightharpoons d^*$ &  $(3,1)$ & $1$ & $\phantom{-}0$ & $\phantom{-} 1$ & $\phantom{-}\frac{2}{3}$  \\[1mm]
  $D_R$ & $a \leftrightharpoons d$  & $( 3,1)$&    $1$ & $\phantom{-}0$ & $- 1$ & $-\frac{1}{3}$   \\[1mm]
  $L_L$ & $c \leftrightharpoons b$ &  $(1,2)$&    $0$ &  $\phantom{-}1$ & $\phantom{-}0$ & $-\frac{1}{2}$ \\[1mm]
  $E_R$ & $c \leftrightharpoons d$ &  $(1,1)$&   $0$ & $\phantom{-}1$ &  $- 1$ & $- 1$ \\[1mm]
 $Q_L$ & $a \leftrightharpoons b$ &  $(3,2)$& $1$ & $\phantom{-}0 $ & $\phantom{-} 0$ & $\phantom{-} \frac{1}{6}$    \\[1mm]
   $N_R$  &  $\phantom{^*}c \leftrightharpoons d^*$  &   $(1,1)$& $0$ & $\phantom{-}1$ & 
$\phantom{-} 1$ & $\phantom{-} 0$ \\ [1mm]
\hline
\hline
\label{table}
\end{tabular}
\end{center}
\end{table}

The covariant derivative for the $U(1)$ fields in the $a, b, c, d$
basis is found to be
\be\el{covderi2} \CD_\mu = \p_\mu - i g'_a \, C_\mu  \,  Q_a   -i  g'_c \, B_\mu \,  Q_c  -i g'_d \, X_\mu \, Q_d \, .\ee
The fields $C_\mu, B_\mu, X_\mu$ are related
to $Y_\mu, Y_\mu{}'$ and $Y_\mu{}''$ by the rotation matrix
\begin{equation}
\CR=
\left(
\begin{array}{ccc}
 C_\theta C_\psi  & -C_\phi S_\psi + S_\phi S_\theta C_\psi  & S_\phi
S_\psi +  C_\phi S_\theta C_\psi  \\
 C_\theta S_\psi  & C_\phi C_\psi +  S_\phi S_\theta S_\psi  & - S_\phi
C_\psi + C_\phi S_\theta S_\psi  \\
 - S_\theta  & S_\phi C_\theta  & C_\phi C_\theta
\end{array}
\right) \,,
\end{equation}
with Euler angles $\theta$, $\psi,$ and $\phi$. Equation~(\ref{covderi2}) can be rewritten in terms of $Y_\mu$, $Y'_\mu$, and
$Y''_\mu$ as follows
\begin{eqnarray}
\CD_\mu & = & \partial_\mu -i Y_\mu \left(-S_\xt g'_d Q_d + C_\theta S_\psi  g'_c  Q_c +  C_\theta C_\psi g'_a Q_a \right) \nonumber \\
 & - & i Y'_\mu \left[ C_\theta S_\phi  g'_d Q_d +\left( C_\phi C_\psi + S_\theta S_\phi S_\psi \right)  g'_c Q_c +  (C_\psi S_\theta S_\phi - C_\phi S_\psi) g'_a Q_a \right] \label{linda} \\
& - & i Y''_\mu \left[ C_\theta C_\phi g'_d Q_d +  \left(-C_\psi S_\phi + C_\phi S_\theta S_\psi \right)  g'_c  Q_c + \left( C_\phi C_\psi S_\theta + S_\phi S_\psi\right) g'_a Q_a \right]   \, .  \nonumber
\end{eqnarray}
Now, by demanding that $Y_\mu$ has the
hypercharge $Q_Y$ given in~\er{hyperchargeY}  we  fix the first column of the rotation matrix $\CR$
\begin{equation}
\bay{c} C_\mu \\  B_\mu \\ X_\mu
\eay = \left(
\begin{array}{lr}
  \phantom{-} Y_\mu \, \frac{1}{6} g_Y /g'_a& \dots \\
 - Y_\mu \, \frac{1}{2} g_Y/g'_c & \dots\\
  \phantom{-} Y_\mu \, \frac{1}{2} g_Y/g'_d & \dots
\end{array}
\right) \, ,
\end{equation}
and we determine the value of the two associated Euler angles
\begin{equation}
\theta = {\rm -arcsin} \left[\frac{1}{2} g_Y/g'_d \right]
\label{theta}
\end{equation}
and
\begin{equation}
\psi = {\rm arcsin}  \left[-\frac{1}{2} g_Y/ (g'_c \, C_\theta) \right] \, .
\label{psi}
\end{equation}
The couplings $g'_c$ and $g'_d$ are related through the orthogonality
condition,
\begin{equation}
 \left(-\frac{1}{2 g' _c} \right)^2  = \frac{1}{g_Y^2} - \left(\frac{c_1}{6g'_a} \right)^2  - \left(\frac{1}{2g'_d}\right)^2  \, ,
\end{equation}
with $g'_a$ fixed by the relation $g_3 (M_s) = \sqrt{6} \, g'_a
(M_s)$.\footnote{Throughout $g_3$ and $g_2$ are the strong and weak
  gauge coupling constants.} In our calculation we take $M_s = 20~{\rm
  TeV}$ as a reference point for running down to 1.8~TeV the $g'_a$
coupling, ignoring mass threshold effects of stringy states. This
yields $g'_a = 0.36$. We have checked that the running of the $g'_a$
coupling does not change significantly for different values of the
string scale.  The third Euler angle $\phi$ and the coupling $g'_d$
will be determined by requiring sufficient suppression to leptons to
accommodate (\ref{cuatro}) and a (pre-cut) production rate $\sigma(pp
\to Z') \times {\cal B} (Z' \to jj)$ in agreement with (\ref{dos}).

\begin{figure}[t]
\begin{center}
\includegraphics[width=0.33\linewidth]{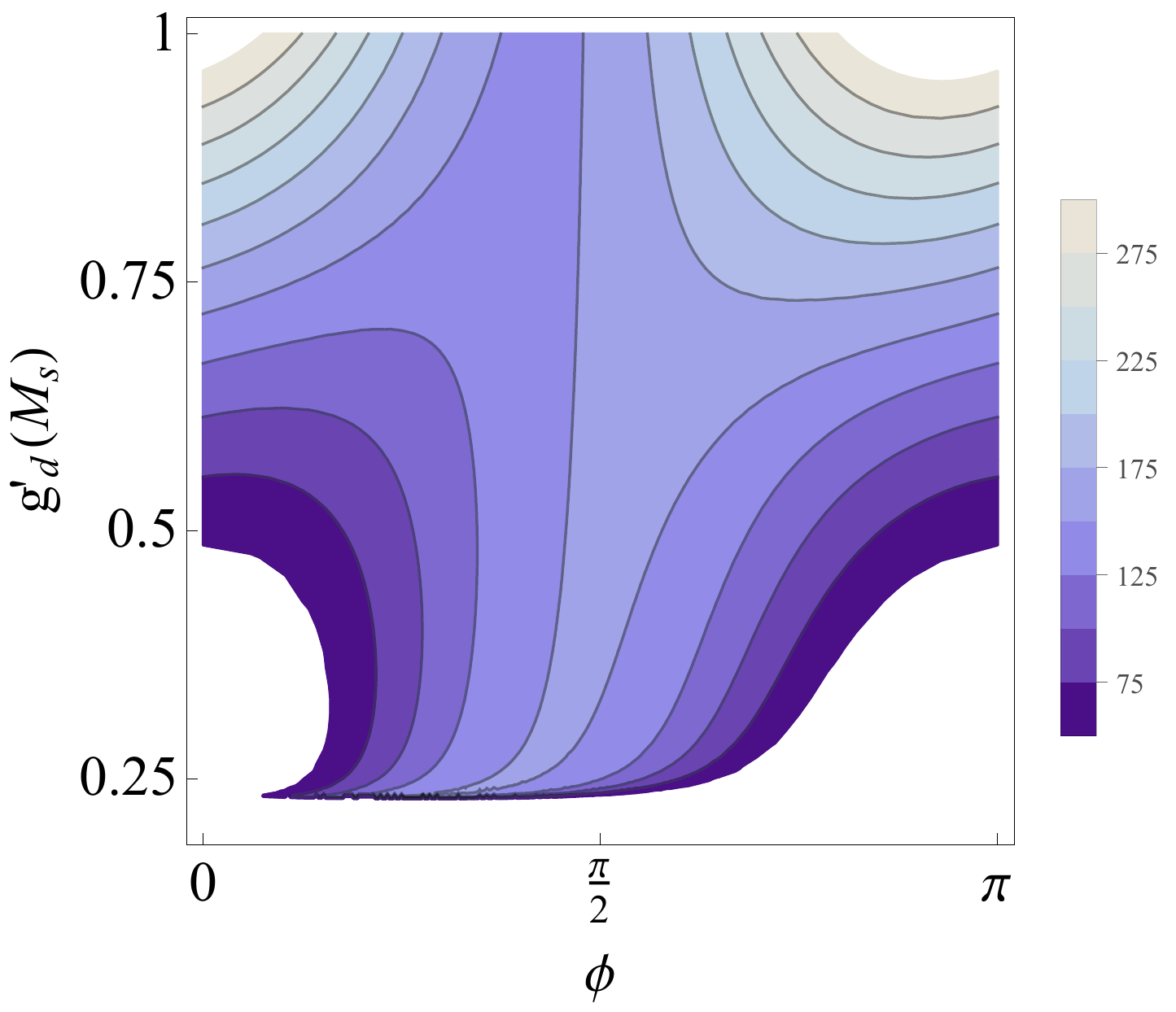}
\includegraphics[width=0.33\linewidth]{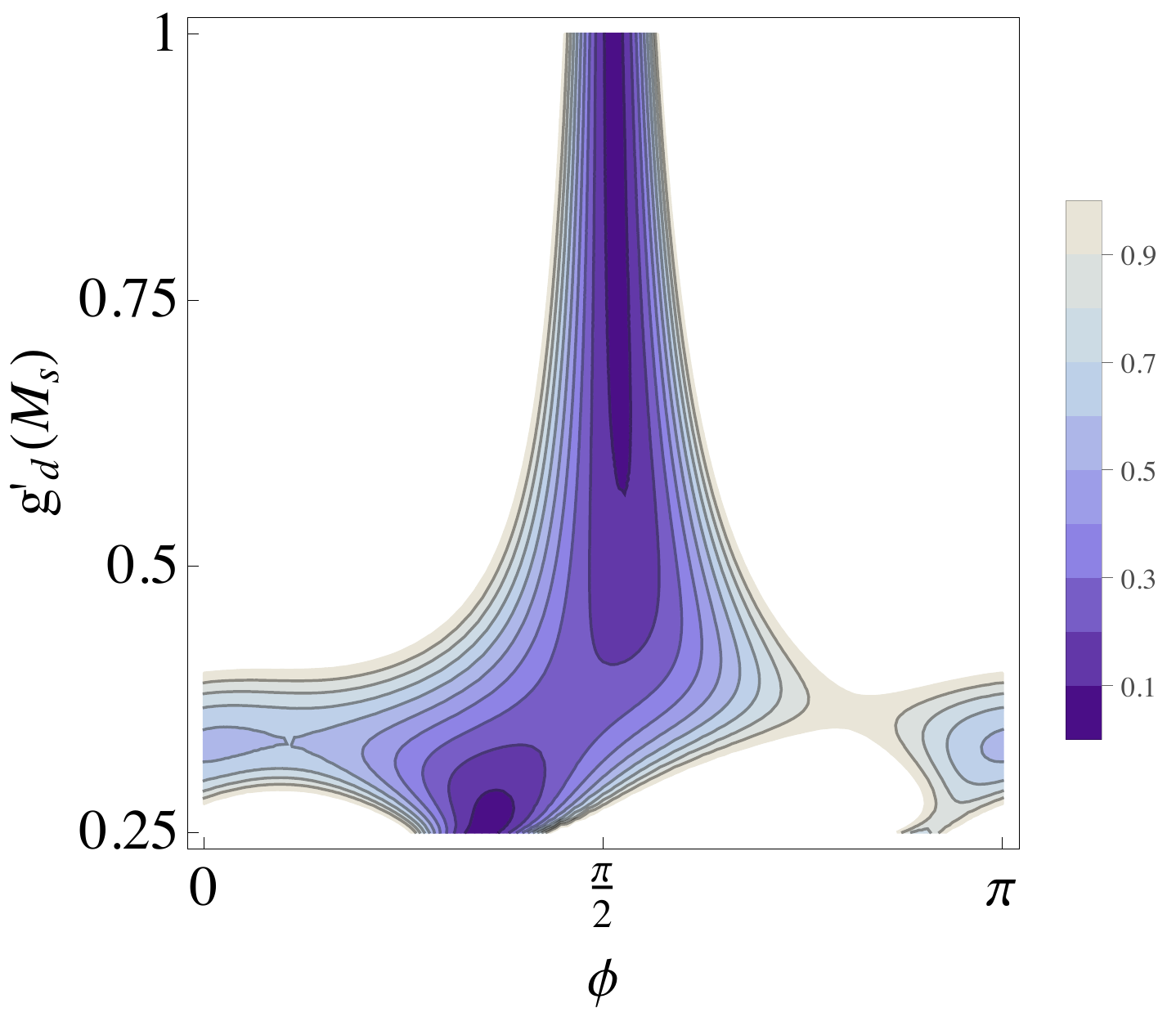}
\includegraphics[width=0.32\linewidth]{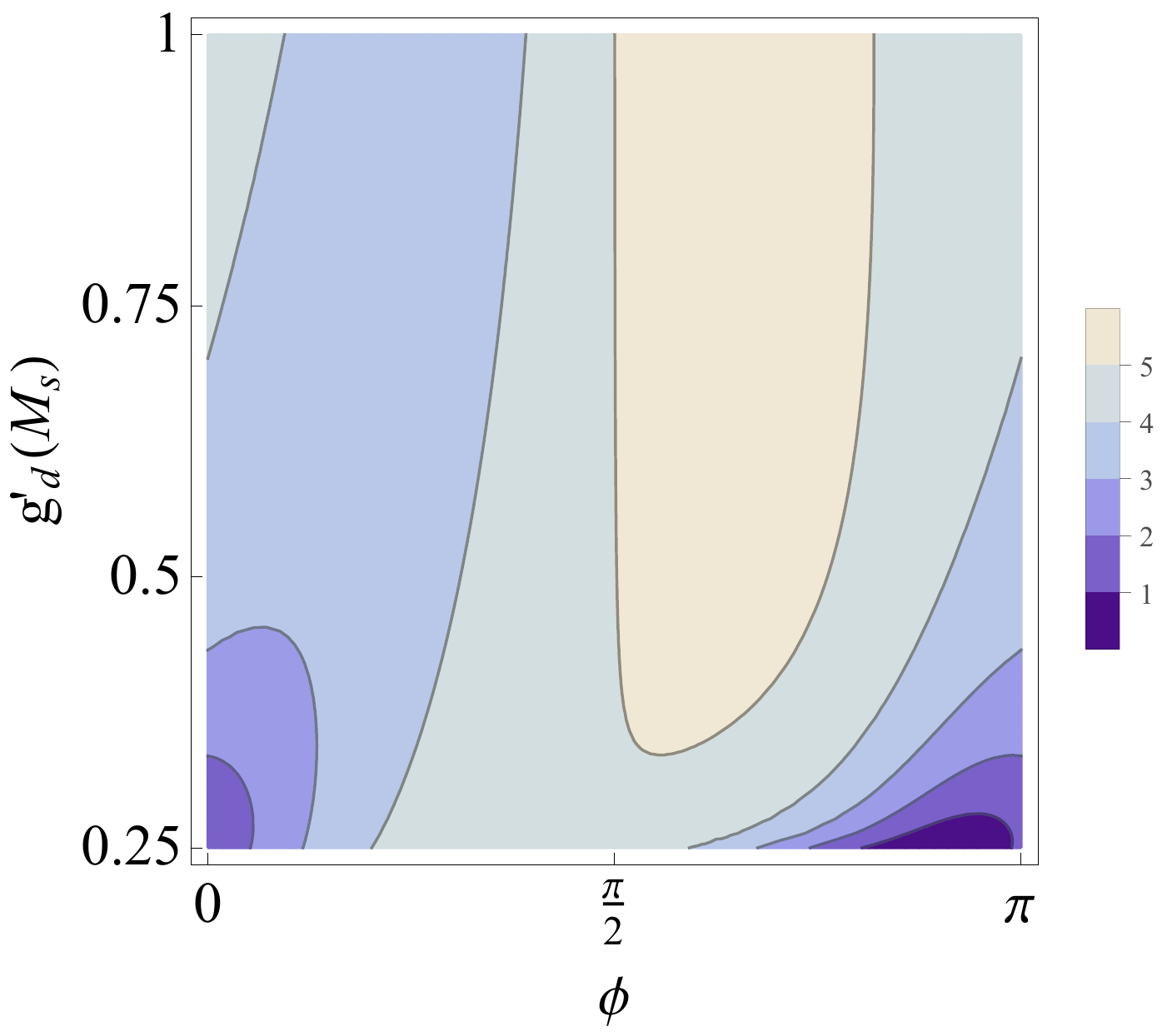}
\end{center}
\vspace{-0.3cm}
\caption[]{Contours of constant cross section $pp \to Z'$ times
  branching into dijet (left), $e^+ e^-$ (middle), and diboson
  (right), for $M_{Z'} \simeq 1.8~{\rm TeV}$ and $\sqrt{s} = 8~{\rm
    TeV}$. The color encoded scales are in  fb.}
\label{fig1}
\end{figure}

\begin{figure}[t]
\begin{center}
\includegraphics[width=0.49\linewidth]{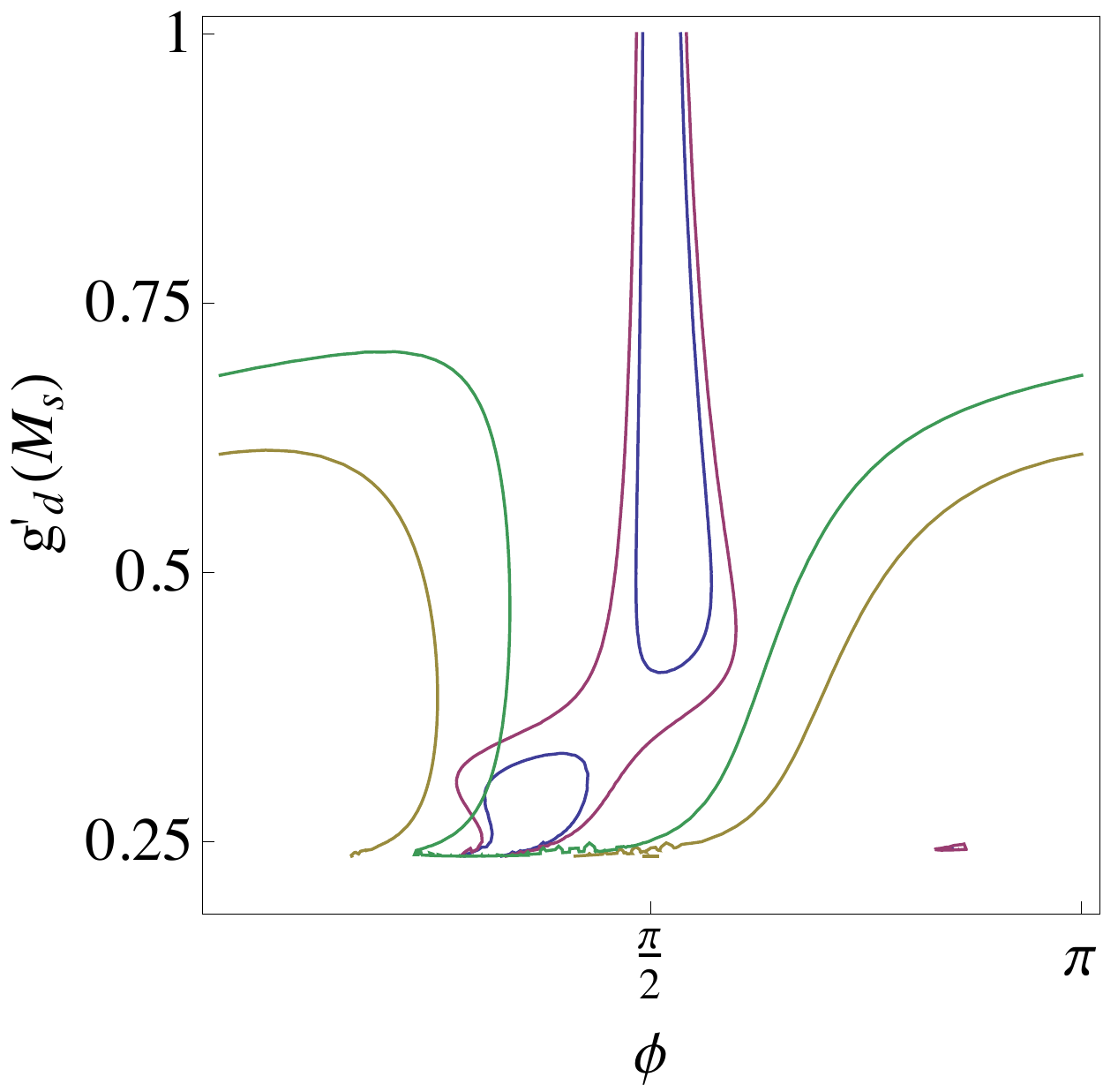}
\includegraphics[width=0.49\linewidth]{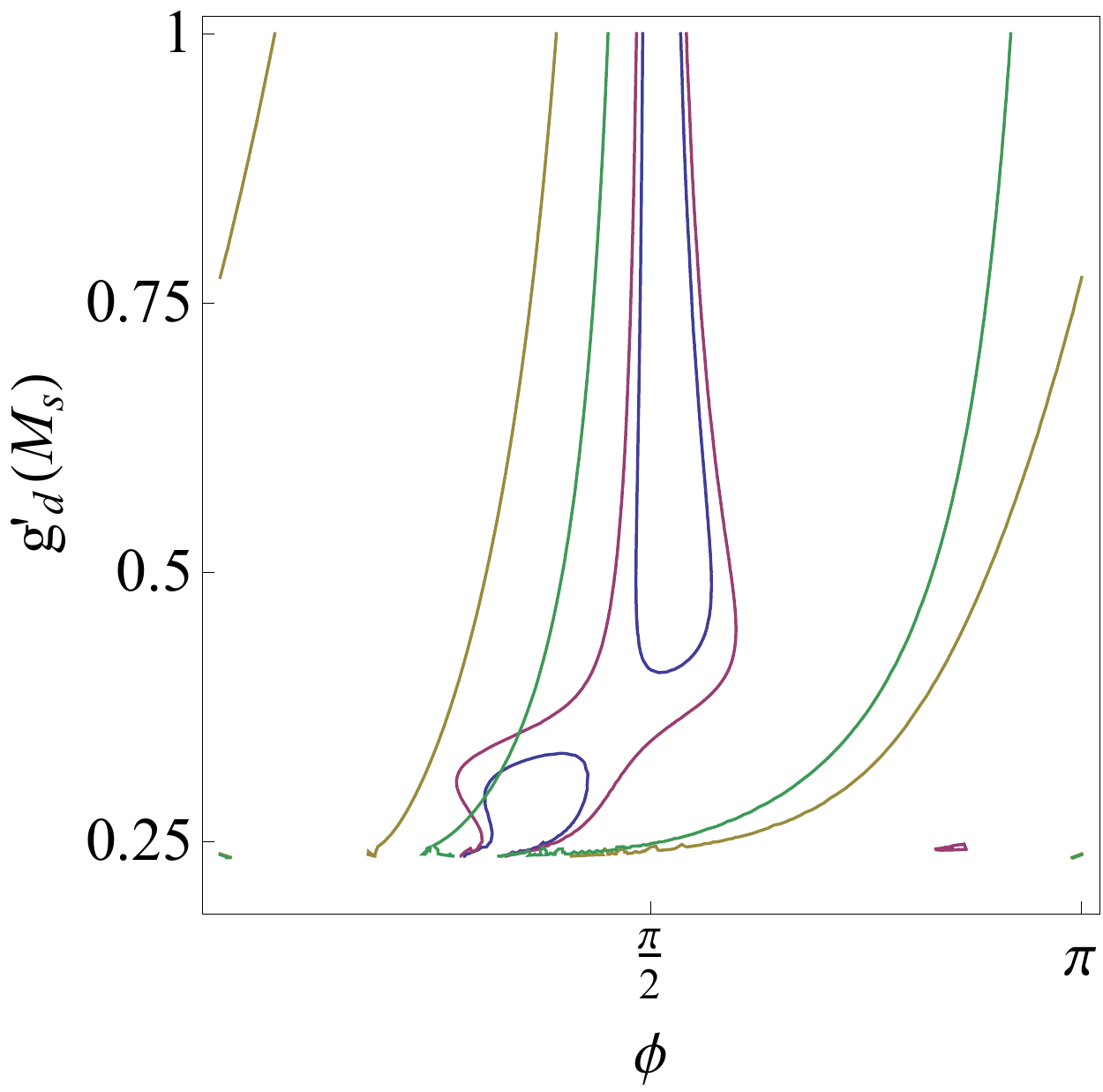}
\end{center}
\vspace{-0.3cm}
\caption[]{Best fit  of cross section contours $pp \to Z'$ times
  branching into dijet/leptons (left) and $pp \to Z'$ times
  branching into dibosons (right), for $M_{Z'} \simeq 1.8~{\rm TeV}$ and $\sqrt{s} = 8~{\rm
    TeV}$. The blue and
  red contours correspond to $\sigma (pp \to Z') \times {\cal B}
  (e^+e^-) = 0.2$ and $0.3~{\rm fb}$, respectively. The yellow and
  green contours on the left correspond to $\sigma (pp \to Z') \times {\cal B}
  (jj) = 91$ and $123~{\rm fb}$, respectively. The yellow and green
  contours on the right correspond to $\sigma (pp \to Z') \times {\cal
    B} (W^+ W^-) =4$ and 4.5~fb, respectively.}
\label{fig2}
\end{figure}

The $f \bar f Z'$ Lagrangian is of the form
\begin{eqnarray}
{\cal L} & = & \frac{1}{2}   \sqrt{g_Y^2 + g_2^2} \ \sum_f \bigg(\epsilon_{f_L} \bar \psi_{f_L} \gamma^\mu \psi_{f_L} +   \epsilon_{f_R} \bar \psi_{f_R} \gamma^\mu \psi_{f_R} \bigg) \, Z'_\mu \, \nonumber \\
& = & \sum_f \bigg((g_{Y'}Q_{Y'})_{f_L} \, \bar \psi_{f_L} \gamma^\mu \psi_{f_L} +  (g_{Y'}Q_{Y'})_{f_R} \bar \psi_{f_R} \gamma^\mu \psi_{f_R} \bigg) \, Z'_\mu \,,
\label{lagrangian}
\end{eqnarray}
where each $\psi_{f_{L \, (R)}}$ is a fermion field with the
corresponding $\gamma^\mu$ matrices of the Dirac algebra, and
$\epsilon_{f_L,f_R} = v_q \pm a_q$, with $v_q$ and $a_q$ the vector
and axial couplings respectively.  From (\ref{linda}) and
(\ref{lagrangian}) we obtain the explicit form of the chiral couplings
in terms of $\phi$ and $g'_d$
\begin{eqnarray}
\epsilon_{u_L} = \epsilon_{d_L} & = & \frac{2}{\sqrt{g_Y^2 + g_2^2}} \, (C_\psi S_\theta S_\psi - C_\phi S_\psi) g'_a \,, \nonumber \\
\epsilon_{u_R} & = &- \frac{2}{\sqrt{g_Y^2 + g_2^2}} \, [C_\theta S_\phi g'_d + (C_\psi S_\theta S_\psi - C_\phi S_\psi) g'_a] \,, \label{couplingphig}\\
\epsilon_{d_R} & = & \frac{2}{\sqrt{g_Y^2 + g_2^2}} \, [C_\theta S_\phi g'_d - (C_\psi S_\theta S_\psi - C_\phi S_\psi) g'_a] \, . \nonumber
\end{eqnarray}
The decay width of $Z' \to f\bar f$ is given by~\cite{Barger:1996kr}
\begin{equation}
\Gamma(Z' \to f \bar f) = \frac{G_F M_Z^2}{6 \pi \sqrt{2}}  N_c C(M_{Z'}^2) M_{Z'} \sqrt{1 -4x} \left[v_f^2 (1+2x) + a_f^2 (1-4x) \right] \, ,
\end{equation}
where $G_F$ is the Fermi coupling constant, $C(M_{Z'}^2) = 1 +
\alpha_s/\pi + 1.409 (\alpha_s/\pi)^2 - 12.77 (\alpha_s/\pi)^3$,
$\alpha_s = \alpha_s(M_{Z'})$ is the strong coupling constant at the
scale $M_{Z'}$, $x = m_f^2/M_{Z'}^2$, and $N_c =3$ or 1 if $f$ is a
quark or a lepton, respectively.  The couplings of the $Z'$ to
the electroweak gauge bosons are model dependent, and are strongly dependent
on the spectrum of the hidden sector.
Following~\cite{Antoniadis:2009ze} we parametrize the model-dependence
of the decay width in terms of two dimensionless coefficients,
\begin{eqnarray}
\Gamma (Z' \to ZZ) &  = &  \frac{c_1^2 \sin^2 \theta_W
  M_{Z'}^3}{192 \pi  M_Z^2} \left( 1 - \frac{4
      M_Z^2}{M_{Z'}^2} \right)^{5/2}  \nonumber \\
& \approx & c_1^2 \ (45~{\rm GeV})
\left(\frac{M_{Z'}}{\rm TeV}\right)^3 + \cdots,
\label{Ignatios-1}
\end{eqnarray}
\begin{eqnarray}
\Gamma (Z' \to W^+ W^-) & = & \frac{c_2^2 M_{Z'}^3}{48 \pi M_W^2}  \left( 1 - \frac{4
      M_W^2}{M_{Z'}^2} \right)^{5/2} \nonumber \\
& \approx&  c_2^2 \ (1.03~{\rm TeV})
    \left(\frac{M_{Z'}}{\rm TeV} \right)^3 + \cdots,  
\label{Ignatios-2}
\end{eqnarray}
\begin{eqnarray}
\Gamma (Z' \to Z\gamma) &  = &  \frac{c_1^2 \cos^2 \theta_W
  M_{Z'}^3}{96 \pi M_Z^2} \left( 1 - \frac{
      M_Z^2}{M_{Z'}^2} \right)^3 \left(1 + \frac{M_Z^2}{M_{Z'}^2}
  \right)  \nonumber \\
& \approx & c_1^2 \ (307~{\rm GeV}) 
  \left(\frac{M_{Z'}}{{\rm TeV}}\right)^3 + \cdots . 
\label{Ignatios-3}
\end{eqnarray}
The $Z'$ production cross section  at the LHC8 is found to be~\cite{Hisano:2015gna} 
\begin{equation}
\sigma (pp \to Z') \simeq 5.2 \left(\frac{2 \Gamma (Z' \to u \bar u) +
    \Gamma (Z' \to d \bar d)}{{\rm GeV} }\right)~{\rm fb} \, .
\end{equation}
Next, we scan the parameter space to obtain agreement with (\ref{uno})
to (\ref{cinco}). In Fig.~\ref{fig1} we show contour plots, in the
$(g'_d, \phi)$ plane, for constant $\sigma(pp \to Z') \times {\cal B}
(Z' \to jj)$, $\sigma(pp \to Z') \times {\cal B} (Z' \to e^+ e^-)$,
and $\sigma(pp \to Z') \times {\cal B} (Z' \to W^+ W^-)$.  To
accommodate (\ref{uno}), (\ref{dos}), and (\ref{cuatro}) the ratio of
branching fractions of electrons to quarks must be minimized subject
to sufficient dijet and diboson production. It is easily seen in
Fig.~\ref{fig2} that $\phi = 0.96$ and $g'_d (M_s) = 0.29$, $c_1 =
0.08$, and $c_2 = 0.02$ yield $\sigma(pp \to Z') \simeq 228~{\rm fb}$,
${\cal B} (Z' \to jj ) \simeq 0.54$, ${\cal B} (Z' \to e^+e^-) \simeq
8.9 \times 10^{-4}$, ${\cal B} (Z' \to WW/ZZ) \simeq 3.4
\times 10^{-2}$, which are consistent with (\ref{uno}), (\ref{dos}),
and (\ref{cuatro}) at the $1\sigma$ level. In addition,  ${\cal B} (Z' \to HZ) \simeq 7.4 \times
10^{-4}$. Thus, the upper limit set by (\ref{cinco})
is also satisfied by our fiducial values of $\phi$, $g'_d$, $c_1$,
and $c_2$.  The chiral couplings of $Z'$ and $Z''$ are given in
Table~\ref{table-ii}. All fields in a given set have a common $g_{Y'} Q_{Y'},\, g_{Y''} Q_{Y''}$ couplings.

\begin{table}
  \caption{Chiral couplings of  $Z'$ and $Z''$ gauge bosons.}
\begin{tabular}{cccc}
\hline
\hline
~~~~~Fields~~~~~ &  ~~~~~$g_{Y'} Q_{Y' }$~~~~~  &  ~~~~~$g_{Y''} Q_{Y''}$~~~~~  \\ \hline
$U_R$ & $-0.24$ & $\phantom{-}0.26$ \\
$D_R$ & $-0.41$ & $\phantom{-}0.02$  \\ 
$L_L$ & $\phantom{-}0.06$ & $\phantom{-}0.23$  \\ 
$E_R$ &   $-0.02$ & $\phantom{-}0.10$ \\ 
$Q_L$  &  $-0.33$ & $\phantom{-}0.14$ \\
$N_R$  & $\phantom{-}0.15$ & $\phantom{-}0.35$ \\
\hline
\hline
\end{tabular} 
\label{table-ii}
\end{table}

The second constraint on the model derives from the mixing of the $Z$
and the $Y'$ through their coupling to the two Higgs doublets. The
criteria we adopt here to define the Higgs charges is to make the
Yukawa couplings ($H_u \bar u q$, $H_d \bar d q$, $H_d \bar e \ell,$
$H_\nu \bar \nu \ell$) invariant under all three
$U(1)$'s~\cite{Antoniadis:2000ena}.  Two ``supersymmetric'' Higgses
$H_u\equiv H_\nu$ and $H_d$ (with charges $Q_a = Q_{c} = 0$, $Q_{d} =
1$, $Q_Y = 1/2$ and $Q_a = Q_{c} = 0$, $Q_{d} = -1$, $Q_Y = -1/2$) are
sufficient to give masses to all the chiral fermions.  Here, $\br H_u
\ke = (^0_{v_u})$, $\br H_d \ke = (^{v_d}_0),$ $v = \sqrt{v_u^2 +
  v_d^2} = 174~{\rm GeV}$, and $\tan \beta \equiv v_u/v_d$.

The last two terms in the covariant derivative
\begin{equation}
\CD_\mu = \p_\mu - i \frac{1}{\sqrt{g_2^2 + g_Y^2}} Z_\mu (g_2^2 T^3 - g_Y^2 Q_Y) -i g_{Y'} Y_\mu{}' Q_{Y'} -i g_{Y''}
Y_\mu{}'' Q_{Y''} ,
\end{equation}
are conveniently written as
\begin{equation}
-i \frac {x_{H_i}}{ v_i} \overline M_Z Y_\mu{}' - i \frac{y_{H_i}} {v_i} \overline M_Z Y_\mu{}'' 
\end{equation}
for each Higgs $H_i$, with $T^3 = \sigma^3/2$, where for the two Higgs doublets
\begin{equation}
x_{H_u} = - x_{H_d} =   1.9 \ \sqrt{{g'_d}^2 -0.032} \ S_\phi  \end{equation}
and
\begin{equation}
y_{H_u} = - y_{H_d}    =  1.9 \ \sqrt{{g'_d}^2 -0.032}
 \ C_\phi \, .
\end{equation}

The Higgs field  kinetic term together with the GS mass
terms  ($-\frac{1}{2} M'^2 Y'_\mu Y'^\mu - \frac{1}{2} M''^2 Y''_\mu
Y''^\mu$) yield the following mass square matrix for the $Z-Z'$ mixing,
$$ \bay{ccc} \overline M_Z^2 & \overline M_Z^2 (x_{H_u} C_\beta^2 - x_{H_d} S_\beta^2) & \overline M_Z^2 (y_{H_u}  C_\beta^2 - y_{H_d} S_\beta^2) \\ \overline M_Z^2 (x_{H_u} C_\beta^2 - x_{H_d} S_\beta^2) &
\overline M_Z^2 (C_\beta^2 x_{H_u}^2 + S_\beta^2 x_{H_d}^2) + M'^2 &
\overline M_Z^2
(C_\beta^2 x_{H_u} y_{H_u} + S_\beta^2  x_{H_d} y_{H_d})  \\
\overline M_Z^2 (y_{H_u} C_\beta^2 - y_{H_d} S_\beta^2) & \overline
M_Z^2 (C_\beta^2 x_{H_u} y_{H_u} + S_\beta^2 x_{H_d} y_{H_d}) &
\overline M_Z^2 (y_{H_u}^2 C_\beta^2 + y_{H_d}^2 S_\beta^2) +
M''^2\eay \,, $$ which does not impose any constraint on the $\tan
\beta$ parameter. We have verified that, for
our fiducial values of $\phi$ and $g'_d$, if  $M_{Z''} \agt M_{Z'}$
the shift of the $Z$ mass would lie within 1 standard deviation
of the experimental value. 

In summary, we have shown that recent results by ATLAS and CMS
searching for heavy gauge bosons decaying into $WW/ZZ$ and $jj$ final
states could be a first hint of string physics. In D-brane string
compactifications the gauge symmetry arises from a product of $U(N)$
groups, guaranteeing extra $U(1)$ gauge bosons in the spectrum. The
weak hypercharge is identified with a linear combination of anomalous
$U(1)$'s which itself is anomaly free. The extra anomalous $U(1)$
gauge bosons generically obtain a St\"uckelberg mass. Under the
assumption of a low string scale, we have shown that the diboson and
dijet excesses can be steered by an anomalous $U(1)$ field with very
small coupling to leptons.  The Drell-Yan bounds are then readily
avoided because of the leptophobic nature of the massive $Z'$ gauge
boson. The resulting loop diagrams, along with tree-level
higher-dimension couplings arising from the GS anomaly cancellation
mechanism, generate an effective vertex that couples the anomalous
$U(1)$ fields to two electroweak gauge bosons. The effective vertex
renders viable the decay of the $Z'$ into $Z$-pairs, which is
necessary to fit the data.  Should the excesses become statistically
significant at the LHC13, the associated $Z\gamma$ topology would
become a signature consistent only with a stringy origin.

\acknowledgments{L.A.A.  is supported by U.S. National Science
  Foundation (NSF) CAREER Award PHY1053663 and by the National
  Aeronautics and Space Administration (NASA) Grant No. NNX13AH52G; he
  thanks the Center for Cosmology and Particle Physics at New York
  University for its hospitality.  H.G. and
  T.R.T. are supported by NSF Grant No. PHY-1314774.  X.H.  is
  supported by the MOST Grant 103-2811-M-003-024. D.L. is partially
  supported by the ERC Advanced Grant “Strings and Gravity”
  (Grant.No. 32004) and by the DFG cluster of excellence ``Origin and
  Structure of the Universe.''  Any opinions, findings, and
  conclusions or recommendations expressed in this material are those
  of the authors and do not necessarily reflect the views of the
  National Science Foundation.}


\begin{thebibliography}{999}

\bibitem{Aad:2015owa} 
  G.~Aad {\it et al.}  [ATLAS Collaboration],
  arXiv:1506.00962 [hep-ex].


\bibitem{Khachatryan:2014hpa} 
  V.~Khachatryan {\it et al.}  [CMS Collaboration],
  JHEP {\bf 1408}, 173 (2014)
  [arXiv:1405.1994 [hep-ex]].



\bibitem{Khachatryan:2014gha} 
  V.~Khachatryan {\it et al.}  [CMS Collaboration],
  JHEP {\bf 1408}, 174 (2014)
  [arXiv:1405.3447 [hep-ex]].






\bibitem{Khachatryan:2015sja} 
  V.~Khachatryan {\it et al.}  [CMS Collaboration],
  Phys.\ Rev.\ D {\bf 91}, no. 5, 052009 (2015)
  [arXiv:1501.04198 [hep-ex]].

\bibitem{Aad:2014aqa} 
  G.~Aad {\it et al.}  [ATLAS Collaboration],
  Phys.\ Rev.\ D {\bf 91}, no. 5, 052007 (2015)
  [arXiv:1407.1376 [hep-ex]].

\bibitem{CMS-14-010}
V.~Khachatryan {\it et al.}  [CMS Collaboration],
CONF-Note CMS-PAS-EXO-14-010 (2015).



\bibitem{Brehmer:2015cia} 
  J.~Brehmer, J.~Hewett, J.~Kopp, T.~Rizzo and J.~Tattersall,
  arXiv:1507.00013 [hep-ph].

\bibitem{Hisano:2015gna} 
  J.~Hisano, N.~Nagata and Y.~Omura,
  arXiv:1506.03931 [hep-ph];
K.~Cheung, W.~Y.~Keung, P.~Y.~Tseng and T.~C.~Yuan,
  arXiv:1506.06064 [hep-ph];
 B.~A.~Dobrescu and Z.~Liu,
  arXiv:1506.06736 [hep-ph];
  Y.~Gao, T.~Ghosh, K.~Sinha and J.~H.~Yu,
  arXiv:1506.07511 [hep-ph];
  Q.~H.~Cao, B.~Yan and D.~M.~Zhang,
  arXiv:1507.00268 [hep-ph];
T.~Abe, R.~Nagai, S.~Okawa and M.~Tanabashi,
  arXiv:1507.01185 [hep-ph];
  J.~Heeck and S.~Patra,
  arXiv:1507.01584 [hep-ph];
B.~C.~Allanach, B.~Gripaios and D.~Sutherland,
  arXiv:1507.01638 [hep-ph];
 T.~Abe, T.~Kitahara and M.~M.~Nojiri,
  arXiv:1507.01681 [hep-ph];
  B.~A.~Dobrescu and Z.~Liu,
  arXiv:1507.01923 [hep-ph];
H.~S.~Fukano, S.~Matsuzaki and K.~Yamawaki,
  arXiv:1507.03428 [hep-ph].
See also~\cite{Brehmer:2015cia}. 

\bibitem{Fukano:2015hga} 
  H.~S.~Fukano, M.~Kurachi, S.~Matsuzaki, K.~Terashi and K.~Yamawaki,
  arXiv:1506.03751 [hep-ph];
A.~Thamm, R.~Torre and A.~Wulzer,
  arXiv:1506.08688 [hep-ph];
A.~Carmona, A.~Delgado, M.~Quiros and J.~Santiago,
  arXiv:1507.01914 [hep-ph];
C.~W.~Chiang, H.~Fukuda, K.~Harigaya, M.~Ibe and T.~T.~Yanagida,
  arXiv:1507.02483 [hep-ph];
G.~Cacciapaglia, A.~Deandrea and M.~Hashimoto,
arXiv:1507.03098 [hep-ph];
V.~Sanz,
  arXiv:1507.03553 [hep-ph].

\bibitem{Alves:2015mua} 
  A.~Alves, A.~Berlin, S.~Profumo and F.~S.~Queiroz,
  arXiv:1506.06767 [hep-ph].


\bibitem{Aguilar-Saavedra:2015rna} 
  J.~A.~Aguilar-Saavedra,
  arXiv:1506.06739 [hep-ph].


\bibitem{Chen:2015xql} 
  C.~H.~Chen and T.~Nomura,
  arXiv:1507.04431 [hep-ph].



\bibitem{Green:1984sg} 
  M.~B.~Green and J.~H.~Schwarz,
  Phys.\ Lett.\ B {\bf 149}, 117 (1984);
  E.~Witten,
  Phys.\ Lett.\ B {\bf 149}, 351 (1984);
 M.~Dine, N.~Seiberg and E.~Witten,
  Nucl.\ Phys.\ B {\bf 289}, 589 (1987);
W.~Lerche, B.~E.~W.~Nilsson, A.~N.~Schellekens and N.~P.~Warner,
Nucl.\ Phys.\ B {\bf 299}, 91 (1988);
L.~E.~Ibanez and F.~Quevedo,
  JHEP {\bf 9910}, 001 (1999)
  [hep-ph/9908305].



\bibitem{Antoniadis:1998ig} 
  I.~Antoniadis, N.~Arkani-Hamed, S.~Dimopoulos and G.~R.~Dvali,
  Phys.\ Lett.\ B {\bf 436}, 257 (1998)
  [hep-ph/9804398].

\bibitem{Blumenhagen:2005mu} 
  R.~Blumenhagen, M.~Cvetic, P.~Langacker and G.~Shiu,
  Ann.\ Rev.\ Nucl.\ Part.\ Sci.\  {\bf 55}, 71 (2005)
  [hep-th/0502005];
  R.~Blumenhagen, B.~Kors, D.~L\"ust and S.~Stieberger,
  Phys.\ Rept.\  {\bf 445}, 1 (2007)
  [hep-th/0610327].


\bibitem{Anchordoqui:2007da}
  L.~A.~Anchordoqui, H.~Goldberg, S.~Nawata and T.~R.~Taylor,
  Phys.\ Rev.\ Lett.\  {\bf 100}, 171603 (2008)
  [arXiv:0712.0386 [hep-ph]];
  L.~A.~Anchordoqui, H.~Goldberg, S.~Nawata and T.~R.~Taylor,
  Phys. Rev. D {\bf 78}, 016005 (2008)
  [arXiv:0804.2013 [hep-ph]].




\bibitem{Lust:2008qc}
  D.~L\"ust, S.~Stieberger and T.~R.~Taylor,
  Nucl.\ Phys.\  B {\bf 808}, 1 (2009)
  [arXiv:0807.3333 [hep-th]];
  L.~A.~Anchordoqui, H.~Goldberg, D.~L\"ust, S.~Nawata, S.~Stieberger and T.~R.~Taylor,
  Phys.\ Rev.\ Lett.\  {\bf 101}, 241803 (2008)
  [arXiv:0808.0497 [hep-ph]];
  L.~A.~Anchordoqui, H.~Goldberg, D.~L\"ust, S.~Nawata, S.~Stieberger and T.~R.~Taylor,
  Nucl.\ Phys.\  B {\bf 821}, 181 (2009)
  [arXiv:0904.3547 [hep-ph]];
L.~A.~Anchordoqui, I. Antoniadis, D. C. Dai, W. Z. Feng, H. Goldberg,
X. Huang, D. L\"ust, D. Stojkovic, and T. R. Taylor,
  Phys.\ Rev.\ D {\bf 90},  066013 (2014)
  [arXiv:1407.8120 [hep-ph]].



\bibitem{Anchordoqui:2011ag} 
  L.~A.~Anchordoqui, H.~Goldberg, X.~Huang, D.~L\"ust and T.~R.~Taylor,
  Phys.\ Lett.\ B {\bf 701}, 224 (2011)
  [arXiv:1104.2302 [hep-ph]];
  L.~A.~Anchordoqui, I.~Antoniadis, H.~Goldberg, X.~Huang, D.~L\"ust and T.~R.~Taylor,
  Phys.\ Rev.\ D {\bf 85}, 086003 (2012)
  [arXiv:1107.4309 [hep-ph]].



\bibitem{Aad:2014cka} 
  G.~Aad {\it et al.} [ATLAS Collaboration],
  Phys.\ Rev.\ D {\bf 90}, no. 5, 052005 (2014)
  [arXiv:1405.4123 [hep-ex]];
  V.~Khachatryan {\it et al.} [CMS Collaboration],
  JHEP {\bf 1504}, 025 (2015)
  [arXiv:1412.6302 [hep-ex]].


\bibitem{Khachatryan:2015bma} 
  V.~Khachatryan {\it et al.} [CMS Collaboration],
  arXiv:1506.01443 [hep-ex];
  G.~Aad {\it et al.} [ATLAS Collaboration],
  Eur.\ Phys.\ J.\ C {\bf 75}, no. 6, 263 (2015)
  [arXiv:1503.08089 [hep-ex]].



\bibitem{Cremades:2003qj} 
  D.~Cremades, L.~E.~Ibanez and F.~Marchesano,
  JHEP {\bf 0307}, 038 (2003)
  [hep-th/0302105].

\bibitem{Ghilencea:2002da} 
  D.~M.~Ghilencea, L.~E.~Ibanez, N.~Irges and F.~Quevedo,
  JHEP {\bf 0208}, 016 (2002)
  [hep-ph/0205083].




\bibitem{Anastasopoulos:2006cz}
  P.~Anastasopoulos, M.~Bianchi, E.~Dudas and E.~Kiritsis,
  JHEP {\bf 0611} (2006) 057
  [hep-th/0605225].
 C.~Coriano, N.~Irges and S.~Morelli,
  JHEP {\bf 0707}, 008 (2007)
  [hep-ph/0701010];
  N.~Irges, C.~Coriano and S.~Morelli,
  Nucl.\ Phys.\ B {\bf 789}, 133 (2008)
  [hep-ph/0703127 [HEP-PH]].


\bibitem{Landau:1948kw} 
  L.~D.~Landau,
  Dokl.\ Akad.\ Nauk Ser.\ Fiz.\  {\bf 60}, 207 (1948);
 C.~N.~Yang,
  Phys.\ Rev.\  {\bf 77}, 242 (1950).





\bibitem{Barger:1996kr}
  V.~D.~Barger, K.~M.~Cheung and P.~Langacker,
  Phys.\ Lett.\  B {\bf 381}, 226 (1996)
  [arXiv:hep-ph/9604298].



\bibitem{Antoniadis:2009ze} 
  I.~Antoniadis, A.~Boyarsky, S.~Espahbodi, O.~Ruchayskiy and J.~D.~Wells,
  Nucl.\ Phys.\ B {\bf 824}, 296 (2010)
  [arXiv:0901.0639 [hep-ph]].
See also,  
  J.~Kumar, A.~Rajaraman and J.~D.~Wells,
  Phys.\ Rev.\ D {\bf 77}, 066011 (2008)
  [arXiv:0707.3488 [hep-ph]];
J.~Bramante, R.~S.~Hundi, J.~Kumar, A.~Rajaraman and D.~Yaylali,
  Phys.\ Rev.\ D {\bf 84}, 115018 (2011)
  [arXiv:1106.3819 [hep-ph]];
J.~Kumar, A.~Rajaraman and D.~Yaylali,
  Phys.\ Rev.\ D {\bf 86}, 115019 (2012)
  [arXiv:1209.5432 [hep-ph]].






\bibitem{Antoniadis:2000ena} 
  I.~Antoniadis, E.~Kiritsis and T.~N.~Tomaras,
  Phys.\ Lett.\ B {\bf 486}, 186 (2000)
  [hep-ph/0004214].









\end{thebibliography}
\end{document}